\documentclass[aps,superscriptaddress]{revtex4-1}
\usepackage{graphicx} 
\usepackage[left=2cm, 
right=2cm, top=2cm, bottom=2cm]{geometry}
\usepackage{amsmath}
\usepackage{upgreek}
\usepackage{amsfonts}
\usepackage{amssymb}
\usepackage{amsthm}
\usepackage{bm}
\usepackage{chemmacros}
\usepackage{xcolor}
\usepackage{siunitx}
\usepackage{gensymb}
\usepackage{url}
\usepackage{textcomp}
\usepackage{setspace}
\usepackage{hyperref}
\hypersetup{
    colorlinks=true,
    linkcolor=blue,
    citecolor=green 
    }
\usepackage{lineno}

\begin{document}
\title{Synchronized DNA sources for unconditionally secure cryptography}
\author{Sandra Jaudou*}
\affiliation{Gulliver CNRS, ESPCI Paris, Université PSL, 75005 Paris, France} 

 \author{H\'el\`ene Gasnier*} 
 \affiliation{IMT Atlantique, Inserm, LaTIM, Brest, France} 

 \author{Elias Boudjella*}
 \affiliation{LIMMS, CNRS-Institute of Industrial Science, The University of Tokyo, 4-6-1 Komaba, Meguro-ku, Tokyo, 153-8505 Japan} 

 \author{Marc Can\`eve}
 \affiliation{IMT Atlantique, Inserm, LaTIM, Brest, France} 

 \author{Victoria Bloquert}
\affiliation{Gulliver CNRS, ESPCI Paris, Université PSL, 75005 Paris, France}

  \author{Vasily Shenshin}
\affiliation{Gulliver CNRS, ESPCI Paris, Université PSL, 75005 Paris, France} 

\author{ Tilio Pilet}
 \affiliation{IMT Atlantique, Inserm, LaTIM, Brest, France}

 \author{Sacha Gaucher}
\affiliation{Gulliver CNRS, ESPCI Paris, Université PSL, 75005 Paris, France} 

\author{Soo Hyeon Kim}
 \affiliation{LIMMS, CNRS-Institute of Industrial Science, The University of Tokyo, 4-6-1 Komaba, Meguro-ku, Tokyo, 153-8505 Japan}
  \affiliation{Institute of Industrial Science, The University of Tokyo, 4-6-1 Komaba, Meguro-ku, Tokyo, 153-8505 Japan}

  \author{Philippe Gaborit}
\affiliation{XLIM, University of Limoges, Limoges, France}

\author{Gouenou Coatrieux}
\affiliation{IMT Atlantique, Inserm, LaTIM, Brest, France}

 \author{Matthieu Labousse}
\affiliation{Gulliver CNRS, ESPCI Paris, Université PSL, 75005 Paris, France}

\author{Anthony Genot}
 \affiliation{LIMMS, CNRS-Institute of Industrial Science, The University of Tokyo, 4-6-1 Komaba, Meguro-ku, Tokyo, 153-8505 Japan}

 \author{Yannick Rondelez}
 \email{yannick.rondelez@espci.psl.eu}
\affiliation{Gulliver CNRS, ESPCI Paris, Université PSL, 75005 Paris, France}
\begin{abstract}
Secure communication is the cornerstone of modern infrastructures, from finance and healthcare to defense and elections, yet achieving unconditional security—resistant to any computational attack—remains a fundamental challenge. The One-Time Pad (OTP), proven by Shannon to offer perfect secrecy, requires a shared random key as long as the message, used only once. However, distributing large keys over long distances has been impractical due to the lack of secure and scalable sharing options. Here, we introduce a DNA-based cryptographic primitive that leverages random pools of synthetic DNA to install a synchronized entropy source between distant parties. Our approach uses duplicated DNA molecules—comprising random index-payload pairs—as a shared secret. These molecules are locally sequenced and digitized to generate a common binary mask for OTP encryption, achieving unconditional security without relying on computational assumptions.
We experimentally demonstrate this protocol between Tokyo and Paris, using in-house nanopore sequencing, generating a shared secret mask of $\sim$ 400 Mb with a residual error rate of $\sim 5\times 10^{-5}$, correctable via Bose–Chaudhuri–Hocquenghem (BCH) codes to achieve the usual overall decryption failure rate of $2^{-128}$. The min-entropy of the binary mask meets the most recent National Institute of Standards and Technology requirements (SP 800-90B), and is comparable to that of approved cryptographic random number generators. Critically, our system can resist two types of adversarial interference through molecular copy-number statistics, providing an additional layer of security reminiscent of Quantum Key Distribution (QKD), but without distance limitations. This work establishes DNA as a scalable entropy source for long-distance OTP, enabling high-throughput and secure communications in sensitive contexts. By bridging molecular biology and cryptography, DNA-based key distribution opens a promising new route toward unconditional security in global communication networks.
\end{abstract}

\maketitle
* Equal contributions

\section{Main text}
\begin{figure}
    \centering
    \includegraphics[width=1\linewidth]{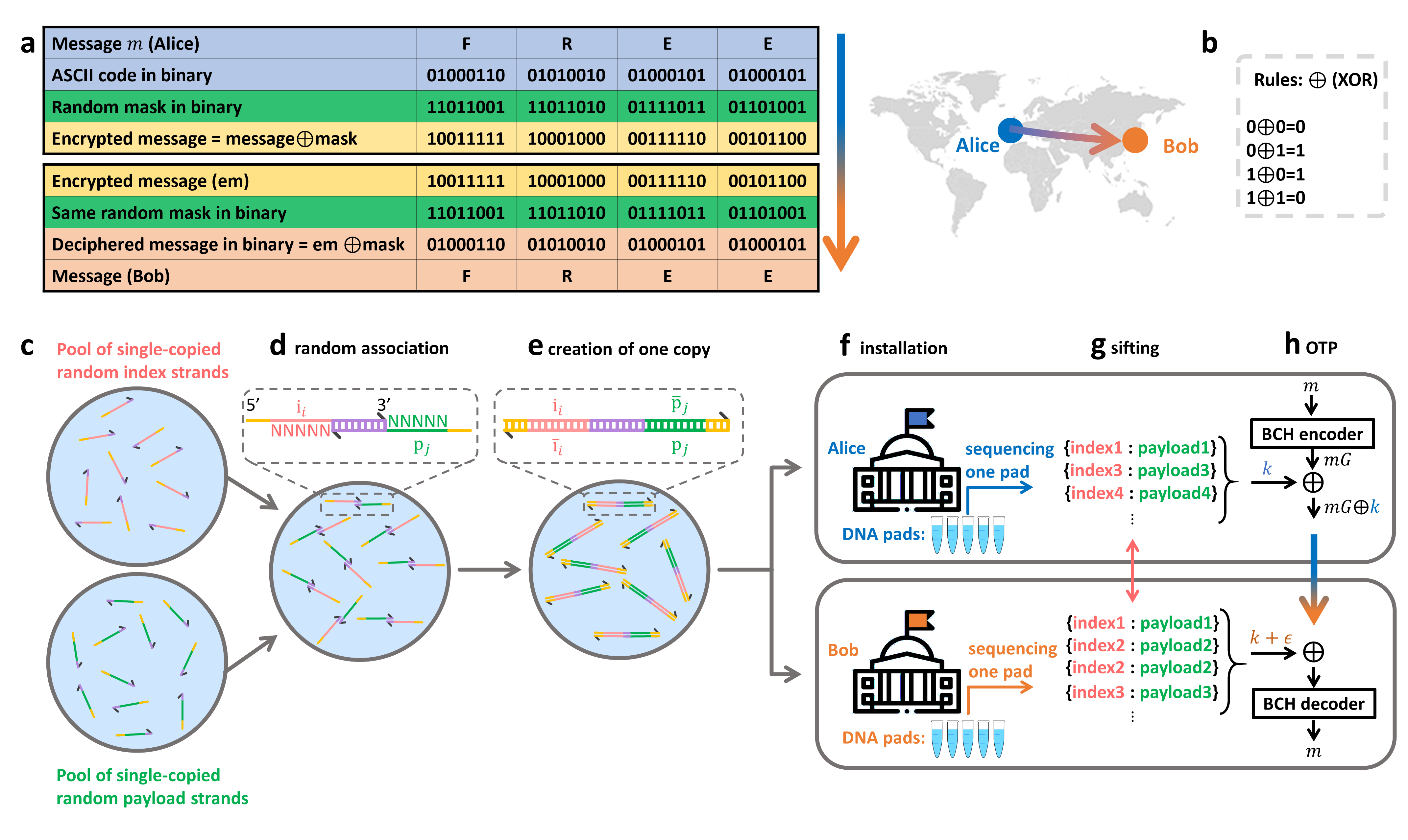}
    \caption{ {\bf DNA-based One-Time-Pad cryptography}. {\bf a}  In OTP cryptography, a message, mapped to binary (e.g. using ASCII-code) is encrypted with a random mask using a bit-by-bit XOR function ({\bf b}). The receiver decodes by applying a bit-by-bit XOR with the same random mask. {\bf c}, Two independent pools containing random DNA strands, each of them being unique, playing the role of index and payload, associate at random ({\bf d}), and are extended over each other by a polymerase to form reverse-complemented duplexes ({\bf e}).  {\bf f}, The pool is optionally amplified  and split into two pads: Alice keeps one and passes the other to Bob. Multiple pads can be duplicated and shared at once, providing synchronized random generation for many future exchanges. {\bf g} Alice and Bob sequence their pad, and publicly share the indices to sift and assemble the corresponding secret payloads into a common binary mask, which they use to communicate via OTP ({\bf h}).}
    \label{fig:Figure1}
\end{figure}

\begin{figure}
    \centering
    \includegraphics[width=1\linewidth]{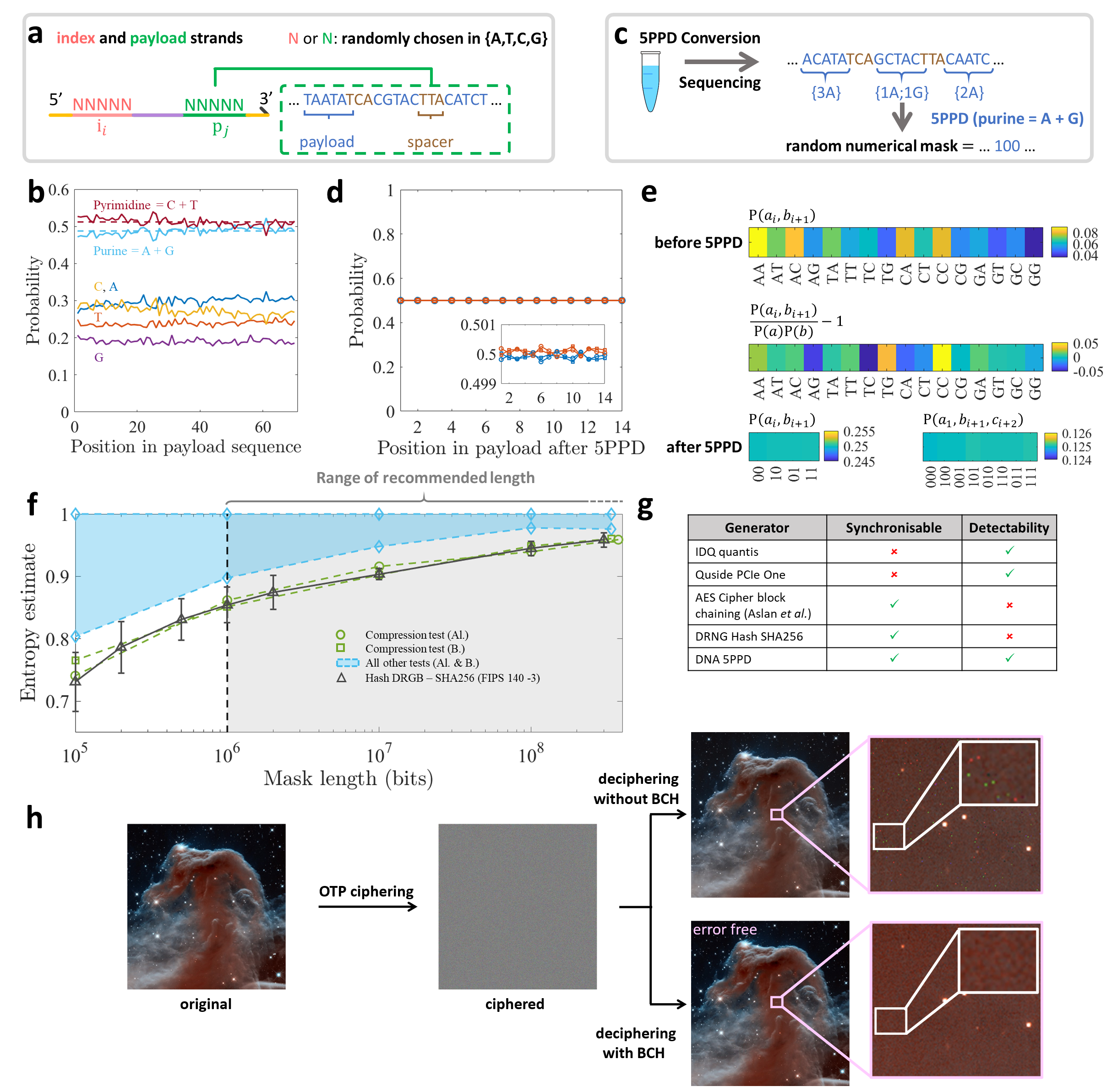}
    \caption{{\bf Generating a shared binary mask using DNA pads}. {\bf a} Index-payload key architecture. {\bf b}, Nucleobase distribution along the payload positions. {\bf c},  Principle of block5 Purine Parity Digitization (5PPD). {\bf d}, Probability of measuring a 1 (red) and 0 (blue) along the binary sequence obtained with a 5PPD of the sequences strands sequenced by Alice (circle) and Bob (square). {\bf e} Pair distribution and correlation  in DNA sequences before 5PPD. Pairs and triplets distribution after 5PPD. {\bf f}, Estimated entropy of DNA binary masks according to the NIST standard 800-90B~\cite{NIST80090B} and comparison with a NIST-approved deterministic RNG~\cite{NISTFIPS} \textcolor{black}{(see Supplementary Tables S4 and S5)}. The standard computes ten entropy estimates and retains the minimum value. The min-entropy is dictated by the compression entropy and all the other estimates are grouped in blue for Alice and Bob sequences. {\bf g} Comparison with commercial RNGs~\cite{Aslan2025}. {\bf h} DNA-OTP ciphering of 2704 $\times$ 2826 image, 130 Mb, of the Horsehead Nebula in the constellation of Orion. Credits: NASA, ESA, and the Hubble Heritage Team (AURA/STScI)~\cite{ESA}).}
    \label{fig:Figure2}
\end{figure}

\begin{figure}
    \centering
    \includegraphics[width=0.8\linewidth]{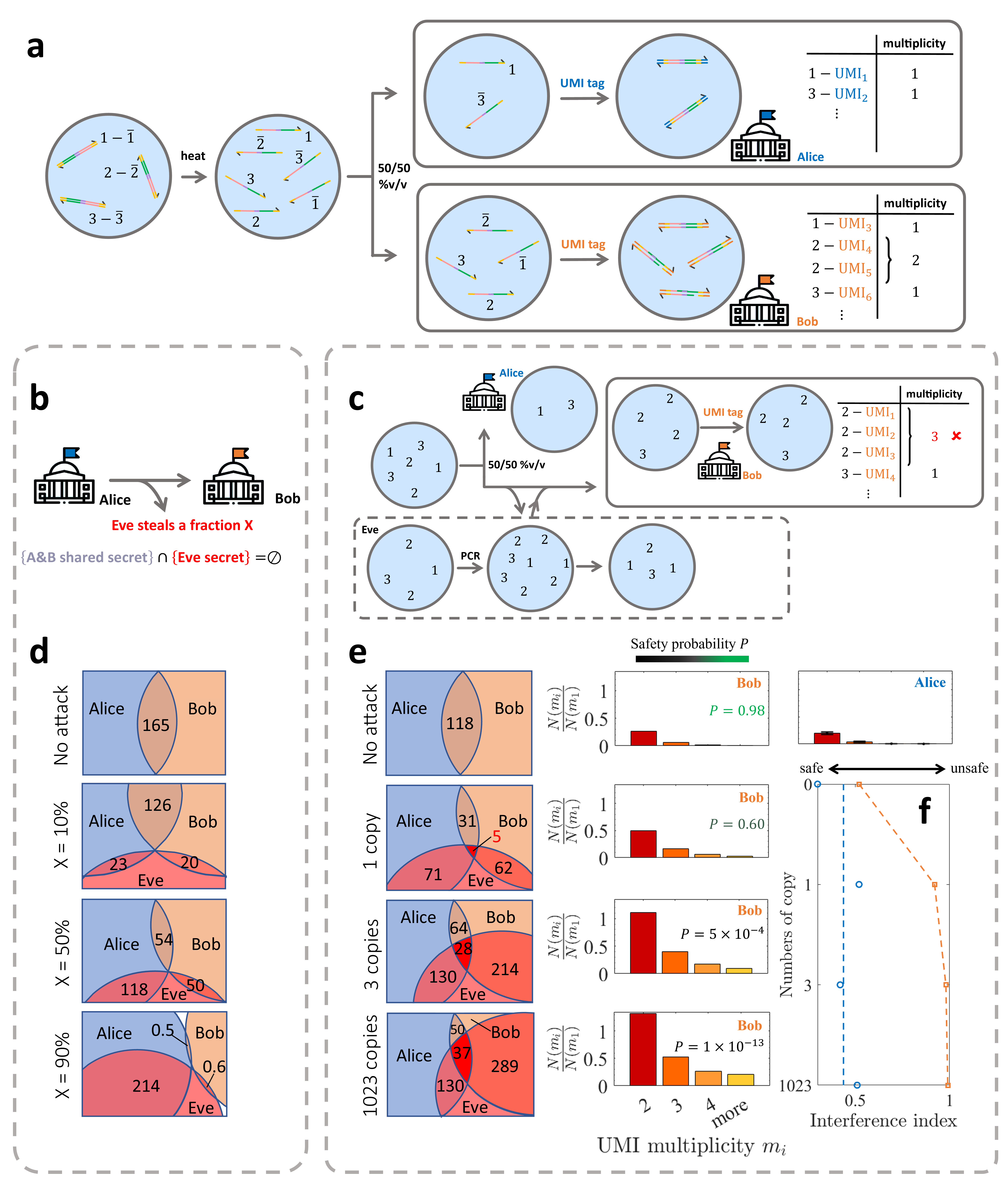}
    \caption{{\bf Securization and simulation of attacks}. {\bf a} Installation of UMI tags to secure the channel. {\bf b} Scenario 1: Eve steals a fraction of the DNA keys within Bob's pad, without replacement. {\bf c} Scenario 2: Eve steals Bob's pad, amplifies the keys by PCR, splits the solution and replace Bob's pool. {\bf d} Ensemble representation of the shared DNA keys for various fractions of theft in scenario 1, showing $\left\lbrace  {\rm Alice } \cap {\rm Bob } \right\rbrace  \cap \left\lbrace  {\rm Eve}  \right\rbrace  =  \O $ in all cases. The diagrams indicate the number of shared keys, in thousands. {\bf e} Ensemble representation for various amplification factors by Eve, and normalized distribution of UMI multiplicity $mi$ in clusters in scenario 2. The four replicates for Alice are shown as a single chart with error bars. The safety probability $P=1-\alpha$ in inset is calculated from the type-I, critical $\alpha$ of $\chi^2$ test of the difference between native (Alices') and intercepted (Bobs') UMI multiplicity distributions. {\bf f}  Interference index defined as $(\sum_{i\geq 2} ( N(m_i)/N(m_1))_{\rm unshared \; keys})/(\sum_{i\geq 2} (N(m_i)/N(m_1))_{\rm shared \; keys})$}
    \label{fig:Figure3}
\end{figure}

{\bf Introduction} During the encryption of a communication, a sender scrambles a plain message using a shared mask which is known only to the sender (Alice) and receiver (Bob), either with a symmetric or an asymmetric protocol. In practice, a combination of the two approaches is often used in cryptography. To communicate securely, Alice and Bob first use a computationally intensive yet robust asymmetric protocol to exchange a small key (such as, {\it e.g.}, Diffie-Hellman~\cite{Diffie1976} or quantum-resistant schemes like ML-KEM~\cite{avanzi2019} or HQC-KEM~\cite{Gaborit2025}). They then cipher and decipher their large messages with their key via an efficient symmetric algorithm, often AES - {\it Advanced Encryption Standards}~\cite{AES}. However, the security of this hybrid approach depends on both the robustness of the asymmetric key exchange protocol and that of the symmetric encryption scheme; ultimately such protocols rely on computational security. With constant increase of computational power available to attackers, security levels must evolve. For instance, in the 80's, keys with 64 bits were considered sufficient, whereas current standards require 128 bits. In practice, it means that data archived 40 years ago with 64 bits hybrid schemes can now be cracked using a modern computer. Basing security guarantees on the computational limits of an attacker introduces a fundamental vulnerability.\\
\\
An alternative to computational key exchange protocols is Quantum Key Distribution (QKD)~\cite{BENNETT20147,Scarani2009,Xu2020}. 
Once a quantum channel is established, Alice and Bob can generate a shared secret key using a physical process whose security is grounded in the fundamental laws of quantum mechanics. Accordingly, QKD resists computational attacks and offers provable security guarantees, including the ability to detect any eavesdropping attempts. This property is particularly significant, as it is not achievable with classical computational or physical key exchange methods, where transmitted data may be copied stealthily, leaving no observable trace. However, fiber-based, twin field or device-independent QKD still remains impractical for sending large keys over distances longer than above $\sim 1000$ km ~\cite{Korzh2015,Lucamarini2018,Pittaluga2021,Wang2022,Zhou2023,Liu2023,Zheng2026,Lu2026}. Although some recent techniques are promising candidates for longer distances in the future~\cite{Gupta2025}, they face the fundamental limitations of quantum repeaters~\cite{Pirandola2017}. Satellite-based QKD~\cite{Liao2018,Liao2017,Li2025} has achieved long distance ground-to-satellite exchange, but this strategy remains limited by weather conditions. Moreover, the short operational time window of low orbit satellites - about 5 min per day - results in a demonstrated record throughput of about $\approx 10^3$ kbit/day~\cite{Li2025}. Finally, the full feasibility of satellite-mediated ground-to-ground QKD has not yet been achieved.\\
\\
Cryptographic schemes providing unconditional security, {\it i.e.}, that remain secure against an adversary with unlimited computational power, do exist and have been known for a long time. In 1949, Shannon demonstrated that the One-Time Pad (OTP) cryptosystem, introduced as early as 1882, offers unconditional security~\cite{Shannon1949}. In OTP cryptography (Fig.~\ref{fig:Figure1}a), Alice and Bob share a secret binary mask. This mask must fulfil the following criteria: be as long as the message to be sent, perfectly random, used only once and destroyed after use. Alice combines the message and the mask with a XOR function. This creates an encrypted message, which can be safely sent over a public channel. At the receiver location, the reverse operation -again a XOR with the shared secret mask- restores the plain message (Fig.~\ref{fig:Figure1}b). Although simple and efficient, OTP encryption poses two challenges, which in practice, strongly limit its applications: first the generation of large masks with high quality of randomness and, second, their secure sharing at two distant locations.\\
\\
The holy grail for secure communication would be to create pairs of unclonable and perfectly random sources that remain synchronized independently of the distance between them. Alice and Bob could then each use their local device to generate large random (but identical) numbers, with which they would communicate with unconditional OTP-based security, a method that remains beyond current reach.\\
\\
Pioneering works~\cite{Meiser2020,Luescher2024} suggest that molecular media, in particular synthetic informational polymers such as DNA, display several interesting features with respect to cryptographic applications~\cite{Wisna2025}. First, the synthesis of random DNA is a simple yet massive source of randomness: lab-scale step-by-step chemical synthesis of a few $1$ mg of DNA with a balanced mixture of the four A, C, G and T nucleotides can produce exabytes ($10^{19}$ bits) of randomness in a few hours. 
Second, unlike many physical sources of entropy \cite{Amer2025}, DNA is natively discrete - canonical DNA is a quaternary alphabet. This discreteness simplifies the mathematical processing of the molecule as an information-carrying medium. 
Third, large amounts of information stored in DNA pools can be duplicated or amplified by autonomous biomolecular operations. This happens in a massively parallel way (typically $10^{12}$ parallel operations per mL) and without the need to extract the information stored at the molecular level. Hence, the milligrams of random DNA mentioned above can be replicated by DNA polymerases and distributed so that two distant parties share exabytes of common, but still unknown, secret. If the molecular strings are properly stored~\cite{Lemaire2025}, the secret can be archived for centuries before retrieval~\cite{Kjaer2022}. 
Fourth, the information contained in large molecular DNA archives can be extracted in a digital form by commercial sequencing machines. These machines currently output terabytes of information per run ~\cite{NHGRI2022}, at rates reaching $\simeq 10^8$ basepairs per second~\cite{Hajieghrari2025}. Some sequencing devices are barely larger than a USB stick, and many are push-and-go operations. 
Lastly, next generation sequencing protocols allow each DNA molecule in a mixture to be uniquely identified and counted~\cite{Konig2010,Kivioja2012}. As we demonstrate below, this direct access to the discrete and stochastic nature of molecular processes can be leveraged to provide an additional level of security.\\
\\
Here, we show that duplicated random DNA pads can be used to synchronize the generation of large random binary numbers at arbitrarily distant locations. We characterize the resulting channel in terms of throughput and the quality of the randomness. We measure the residual mask reconciliation error around $5\times 10^{-5}$, which can be compensated by an error correction code with limited overhead. Using standard lab equipment and a cheap commercial sequencer, we demonstrate that a DNA-based OTP process is currently compatible with intercontinental communication at a rate reaching $10^8 - 10^9$ bits per run and a decryption failure rate below $2^{-128}$. Once the sources are installed, key generation happens without transfer of confidential information, limiting interception options. Still, we show that the channel can be further secured against attacks. We experimentally simulate adversarial interference in two different scenarios and provide statistical measures to detect interception.\\
\\
{\bf Installation of synchronized random sources.}
The creation of duplicated random DNA pads uses three stochastic steps \textcolor{black}{(Methods and supplementary note 1)}. First, Alice orders the synthesis of two pools of partially random DNA oligonucleotides, called index strands and payload strands, from a commercial manufacturer (Fig.~\ref{fig:Figure1}c). During the independent syntheses of the random domain of these strands, random bases are selected from among A, C, G, T. A standard order provides about $\sim 4$ nmol of oligonucleotides, or about $10^{15}$ unique strands, and much larger syntheses are possible. Second, Alice dilutes the strands to about 100 nM and randomly assemble around $10^{12}$ index strands with an equivalent number of payload strands (Fig.~\ref{fig:Figure1}d). A polymerase extends the two oligonucleotides over each other to generate double-stranded index-payload DNA duplexes, which we will call double-stranded DNA keys (Fig.~\ref{fig:Figure1}e). 
Third, a defined number of DNA keys, on the order of ~$10^{6}$-$10^{9}$ molecules, is randomly sampled by taking an aliquot from this pool. Each of these step adds a layer of discrete randomness and contribute to the security of the channel. For example, even if the pool of payload strands was not fully random, or the DNA provider is not fully trusted, it would be impossible to guess which payload strand associated with a particular index strand, or which of these combinations were actually sampled in step three. Importantly, in the resulting DNA key pool, sequence information exists in exactly two copies, in the form of two reverse-complemented DNA molecules. Then, the aliquot is physically partitioned, with or without further amplification, which allows to share information between two pads. Alice keeps one and sends the second to Bob (Fig.~\ref{fig:Figure1}f). The process can be parallelized and repeated to generate multiple duplicated DNA pads, at negligible cost and adaptable capacity \textcolor{black}{(Supplementary note 2)}. After that stage, the synchronized source is installed  and all communications can happen on public channels. \\
\\
{\bf Authentication and creation of a shared secret}. When Alice wants to send a message, she selects a pad, informs Bob, and both enter the process of generating a common random mask. (Fig.~\ref{fig:Figure1}g). Each party independently sequences its pad using standard protocols. The sequencing machines report a list of DNA key sequences, in the form of millions to billions of independent index-payload associations. Alice and Bob's sets overlap, but, because of statistical sampling and biomolecular or sequencing errors, they do not necessarily fully coincide. Bob then publicly sends to Alice his list of index sequences while keeping the associated payloads secret, and Alice compares them to her own list. As the diversity of indices scales exponentially with index length,  Alice can now authenticate Bob with a high level of confidence \textcolor{black}{(Supplementary note 3)}. Then, she computes the intersection between the two sets, decides of a specific ({\it e.g.}, random) index ordering, and publicly sends back that list to Bob. The corresponding ordered payloads then form the shared secret between Alice and Bob, which they convert to a binary OTP mask (Fig.~\ref{fig:Figure1}h). This is equivalent to a sifting stage in QKD. Importantly, no information concerning the payload sequences was exchanged in the process.\\
\\
We localized Alice in Paris and Bob in Tokyo and experimentally tested the full biomolecular protocol. For installation, Alice obtained degenerate oligonucleotides pools from IDT (Integrated DNA Technologies),  assembled, amplified by PCR and split the sample in duplicated pads containing approximately $30\times 10^6$ unique DNA keys. The random parts of the DNA keys were composed of 14 domains of length $n=5$, separated by spacer sequences identical in all strands (Fig.~\ref{fig:Figure2}a), hence a combinatorial space of more than $2^{140}\approx10^{42}$ possible DNA keys. This design was selected to facilitate alignment and digitization of the keys (see below). A pad was sent to Tokyo and stored. For communication the two pads were independently sequenced using nanopore technology on local P2 Solo machines, a miniaturized sequencer with a footprint of just 15$\times$11$\times$9 ${\rm cm}$. After quality filtering and aligning, the two datasets were clustered, and consensus sequences were extracted for each cluster, along with cluster size and quality metric. These metrics were used for a final filtering stage, after which Alice and Bob retained 26 586 748 and 27 915 041 high-quality DNA key sequences, respectively. When exchanging their list of indices, they found an overlap of 22 603 540 exact correspondences \textcolor{black}{(Supplementary Table S1)}.\\
\\
{\bf Generating the binary key and assessing the quality of the randomness}. The simplest approach to binarizing a DNA sequence operates at the nucleobase level, employing a canonical quaternary code ({\it e.g.}, A = 00, G = 01, C = 10, T = 11). This encoding scheme maximizes the amount of digital information that can be extracted from DNA—up to 2 bits per base in theory, but closer to 1.83 bits per base once experimental constraints are considered ~\cite{Erlich2017}. However, it is not suitable for generating random bits from synthetic DNA due to its sensitivity to biases and correlations commonly associated with degenerate DNA synthesis~\cite{Meiser2020}. For example, Fig.~\ref{fig:Figure2}b shows the unbalanced distribution of the four nucleobases with a gradual drift along the chemical synthesis direction, observed in Bob's filtered consensus set. In addition, we observe pairwise correlations (Fig.~\ref{fig:Figure2}e) up to a length of 5 \textcolor{black}{(Supplementary note 4a)}. Consequently, standard debiasing approaches, such the von Neumann protocol~\cite{Meiser2020} which requires independence of the bits, cannot be directly applied to DNA sequences.
Here, we level out position-dependent representation biases, and average over spatial correlations along the polymer chain via blockwise binarization of DNA key sequences \textcolor{black}{(Supplementary note 4c)}. Among this family of functions which compromise between randomness quality and throughput, we selected the block-5 Purine Parity Digitization (5PPD), which counts modulo 2 the number of purines in each block of 5 degenerate bases (Fig.~\ref{fig:Figure2}c). The bits are then concatenated column-wise, generating the binary mask (Fig.~\ref{fig:Figure2}d,e).\\
\\
The randomness of a binary mask is an essential feature of a secure OTP protocol, and its quality must adhere to established cryptographic standards. Here we follow the latest entropy estimation guidelines~\cite{NIST80090B,Saarinen2022}. Furthermore, we select the most conservative scenario by using the min-entropy metric, which is the minimal values obtained over 10 different entropy tests (Fig.~\ref{fig:Figure2}f).  
For our experimental demonstration, applying 5PPD results in a shared binary mask of $316$~Mb. We measured the min-entropy for subsequences of various length (Fig.~\ref{fig:Figure2}f). For the full-length mask, we obtained min-entropy values of 0.9588 from Alice's side and 0.9604 from Bob. Irrespective of the mask length used, these values are on par with the ones produced by a numerical Random Number Generator (RNG) approved by the Standards FIPS 140-3~\cite{NISTFIPS}), {\it e.g.,} Hash-DRBG-SHA256, which applies to all sensitive communications among U.S. federal agencies, and with the ones produced by other commercial RNGs (Fig.~\ref{fig:Figure2}g and comparative table in~\cite{Vrana2025}).\\
\\
{\bf Message sending and correction of residual errors}. To avoid any risk of electronic leakage, sequencing, authentication and binary mask generation are performed at the last moment, using a local hardware. The residual error between the masks is treated via a standard layer of error-correcting code. We selected the Bose–Chaudhuri–Hocquenghem (BCH) cyclic code \textcolor{black}{(Supplementary note 5)}, which is widely accepted for correcting random binary errors~\cite{Hocquenghem,BOSE196068}. To set the BCH code parameters, Alice needs an estimate of the error rate of the  channel -which conceivably may vary with pad storage or experimental conditions. Alice and Bob thus publicly share the 5PPD of an additional random stretch included in the index strand. In the experimental demonstration the two binary differed at 4189/157401800 positions, giving an estimated error rate of  $3\times 10^{-5}$ (the actual error on the whole shared binary mask was measured at $\approx 5 \times 10^{-5}$).
Alice then adjusts the parameters of BCH error correction code such that the probability that Bob cannot reconstruct error-free is lower than the standard cryptographic decryption failure rate of $2^{-128}$, and sends the message. To comply to the OTP requirement, immediately after transmission and decoding, all traces of the binary mask are erased on both side. This includes residual DNA in the sequencing chip or experimental waste (such as liquids or contaminated surfaces) which is degraded chemically using standard lab procedures. We tested the full OTP protocol between Paris and Tokyo by OTP-encrypting, sending and deciphering a large color image using the experimentally generated random masks (Fig.~\ref{fig:Figure2}h and \textcolor{black}{(Supplementary Files 1 for high-resolution pictures)}.\\
\\
 {\bf Securing the DNA-synchronized channel}.  In DNA-OTP, the random DNA keys are assembled directly by Alice and remain unknown to anybody until sequencing. Thanks to the exceptional information density and stability of DNA, DNA pads are extremely compact and can be easily concealed, transported using physical security measures and securely stored for long times. Accordingly, pad transport can be extremely infrequent, or even happen only at installation. Still, it is conceivable that an attacker (Eve) can get access to a DNA pad during storage.  We envision two main scenarios for such an attack (among possibly other options). First, Eve could withdraw a fraction of Bob's pad and sequence it independently, expecting that the partial material loss will go unnoticed. Second, with more time and resources, Eve could steal a full pad, PCR-amplify it, split the amplified solution in two, replace Bob's share and keep the rest for sequencing. We show below that a simple procedure, based on the molecular properties of random DNA pools, is available to resist these two types of attacks.\\
 \\
In this secure version of the protocol, Alice prepares the DNA pads by thermal denaturation and splitting, without PCR preamplification (Fig~\ref{fig:Figure3}a); the stochastic partitioning process thus applies to molecules present in exactly two copies (one direct and one reverse-complemented). In addition, Alice and Bob add a step in their sequencing protocol, where they initially mark each molecule in the pad with a small Unique Molecular Identifier (UMI, Fig.~\ref{fig:Figure3}a). Such identifiers are generally made of a short stretch of random DNA and are commonly used in Next Generation Sequencing workflows~\cite{Stegle2015,Yeom2023}. Once these identifiers are covalently attached to the individual DNA chains, it becomes possible to use the deep sequencing data to unambiguously access the molecular count of each key in the original sample, regardless of the steps and biases that occur during sample processing.  Because Alice's pad was prepared by partitioning a 2-copy sample, she theoretically expects either two UMIs per cluster (when she received both direct and reverse-complemented chains of a given key) or a single UMI (in the case where she received only one of the two chains, that is, among the shared set). By increasing the copy number and introducing an additional partitioning, Eve's copy-and-replace attack will alter the molecular count statistics and become noticeable.\\
\\
To test this concept, we experimentally prepared 10 denaturated DNA pad pairs, from a sampled diversity of approximately 2 million DNA keys each, and simulated Eve's attacks under the two scenarios above, with various intensities (Fig.~\ref{fig:Figure3}b-c). Alice and Bob then entered the key sequencing stage as before, except for the addition of the UMI-tagging preliminary step~\textcolor{black}{(Supplementary note 6)}. After sequencing, Alice (or Bob) groups the reads by their index-payload content, and counts the number of different UMI associated with each of these clusters. As expected, the simple partial theft by Eve resulted in a strictly null tripartite shared set in all cases (\ref{fig:Figure3}d), meaning that the channel remains safe. With a copy-and-replace attack on Bob's pad, Eve could get access to a part of the shared secret, but the statistical analysis of UMI counts within Bob's clusters was clearly affected by the interference (Fig.~\ref{fig:Figure3}e-f). Due to some imperfections in the biomolecular operations, some clusters with multiplicities greater than 2 were observed even in uncompromised samples. However, internal renormalization between the shared and nonshared set provides a very sensitive "interference index". This index reacted even to the most conservative one-copy  attack, which only provided Eve with $15\%$ of the shared secret, a fraction that would be easily mitigated via standard privacy amplification techniques~\cite{Bittau2017}. The interference is also noticable in the corresponding PCR amplification curves \textcolor{black}{(Supplementary note 7)}. \\
\\ 
\textbf{Discussion} Our work introduces a novel paradigm in secure communication by leveraging synthetic DNA as a medium for generating large shared cryptographic keys, combining the proved security of OTP with the scalability and original properties of random DNA polymers.\\
\\
At the heart of this approach lie the unique features of DNA, and its associated biotechnological tools. We demonstrate that DNA-based randomness—arising from the statistical incorporation of nucleotides during chemical synthesis; from the inherent stochasticity of biochemical reactions in the assembly of index-payload DNA keys; and from sampling in high-diversity pools—enables the safe generation of high-quality cryptographic masks. Sharability leverage the double helical pairing of DNA molecules, the same property that enable biological heredity. Security rest on limited attack options, and also exploits the discrete nature of molecular pools, where information can exist on molecules with small -possibly single- copy number. When only a single pair of direct and reverse-complement is present, tripartite sharing is naturally forbidden. Attacks then necessarily involve a molecular-level copying process, which may leave scars, such as detectable anomalies in copy-number statistics. Additional security could be provided by converting the DNA to a non-amplifiable—but still sequencable— informational polymer~\cite{Lutz}, as recently demonstrated~\cite{Kokoris2025}. \\
\\
Compared to more explored alternatives for secure key distribution \cite{BENNETT20147,Scarani2009,Xu2020,Korzh2015,Lucamarini2018,Pittaluga2021,Wang2022,Zhou2023,Liu2023,Zheng2026,Lu2026}, a critical advantage of DNA-based system lies in the capacity to generate synchronized random numbers across very distant locations. While physical transport is required for initial installation, mask generation itself happens without material or photon exchange, only public information is transmitted on a classical channel. The high density and long-term stability of DNA allow for exceptionally infrequent installation: a single gram of DNA pads could support petabytes of unconditionally secure transmissions over extended periods.  Assuming a one-hour hands-on time to process a pad, we estimate the throughput at $\simeq 10^{5}$ bit/s with standard equipment and sequencer, a value that compares favorably with KQD~\cite{Wang2022,Zhou2023,Liu2023,Zheng2026,Lu2026}. Overall, while QKD is grounded in quantum principles but faces challenges in distance and scalability, DNA—and more generally molecular—key generation opens a promising new avenue for cryptographers to explore its operational characteristics~\cite{walter2026}.\\
\\
Because of its density and stability, DNA is also actively explored as a medium for massive digital data storage~\cite{Church2012}. Beside supporting OTP encryption, DNA-based protocols  could be adapted to secure long-term data archiving, where sensitive information stored in DNA databases could only be deciphered by the owner of a matching DNA key.\\
\\
Despite its promises, several challenges must be addressed to fully realize the potential of DNA in OTP or other cryptographic applications. Latency and cost \textcolor{black}{(Supplementary note 2)} remain key considerations, and may be limiting for high-bandwidth communications. Future work could explore faster sequencing technologies, which promise large throughput increase and exponential cost decrease~\cite{Hajieghrari2025}. Combined with automated and packaged workflows for DNA archives manipulation~\cite{Takahashi2019}, this could make DNA-OTP more accessible. Finally, standardization and interoperability are critical for real-world deployment, requiring the establishment of standardized protocols for DNA pad generation, usage, and destruction.These development may however soon open the widespread application of molecular randomness in securing sensitive communications, financial transactions or archiving~\cite{Amer2025}.

\section{Methods}%
{\bf Oligonucleotide sequence design} Index and payload sequences were designed with 14 regions containing 5 random nucleotides (N-blocks) separated by 6-nucleotide defined spacers, summing to 197 and 195 bases, respectively. The 3' end of both oligos are cross-complementary sequences for annealing and extension. The 5' end domains are primer binding site for PCR amplification. The spacers serve two roles: they ease alignment by providing local alignment marks and insulate the variable regions from each other during synthesis, PCR, and sequencing. The dsDNA keys assembly principle is based on the complementarity of 3’ end of these two sequences (index strand and payload strand). Hence, the index strand, possess at 3’ end a 27-nucleotide long region complementary to the 3’ end of the payload strand. Oligonucleotides were ordered at Integrated DNA Technology (IDT) and their sequences are available \textcolor{black}{in Supplementary File 3}. \\
\\
{\bf Generating double-stranded Index-payload DNA keys (dsDNA keys).} Index strands and payload strands were annealed at 100 nM and extended in a 50 $\upmu$L reaction mix containing final concentrations of 1 $\times$ Q5\textregistered{} reaction buffer (NEB, M0493), 200 $\upmu$M dNTP (NEB, N0447), 0.2$\times$ EvaGreen (BIOTIUM, 31000-T), 1\% rAlbumin (NEB, B9200). The following protocol was run on CFX96 Touch Real-Time PCR detection system (BioRad): 25$^{\circ}$C for 10 sec and signal acquisition, 98$^{\circ}$C for 30 sec, 97$^{\circ}$C to 60$^{\circ}$C at -0.5$^{\circ}$C per min, 60$^{\circ}$C for 40 sec and addition of 1\% Q5\textregistered{}  Hot Start High-Fidelity DNA polymerase (NEB, M0493) at this point, 60$^{\circ}$C to 72$^{\circ}$C at +0.5$^{\circ}$C per min and a final extension at 72$^{\circ}$C for 10 min. \\
\\
Post annealing and extension, dsDNA keys, expected at 365 bases long, were purified using SPRIselect beads (Beckman Coulter France, B23318) with a beads-to-sample ratio of 1$\times$, following manufacturer’s recommendation, except that 85\% EtOH was used. dsDNA keys were then quantified using Qubit double-stranded DNA High Sensitivity kit.\\
\\
To estimate the number of dsDNA keys in the sample post purification, which at this stage correspond to the pool’s diversity, an electrophoresis was run on a 4200 TapeStation System (Agilent) using High Sentitivity D1000 reagents and ScreenTape, following manufacturer’s instructions. We measured 60.5 pg of DNA per $\upmu$L for the 365 bp peak, corresponding to 1.5$\times$10$^8$ dsDNA keys per $\upmu$L.\\
\\
{\bf Synchronized random number generation using duplicated DNA pads between ESPCI Paris and the University of Tokyo.} \\
\\
{\it Double stranded DNA keys bottlenecking.} Purified dsDNA keys (quantified at 1.5$\times$10$^8$ molecules per $\upmu$L) were diluted and sampled to obtain approximately 30 million of molecules in 2 $\upmu$L. \\
\\
{\it Amplifying dsDNA keys.} To amplify dsDNA key sample, a PCR was run using a mix containing 1$\times$ Q5 reaction buffer (NEB, M0493), 200 $\upmu$M dNTP (NEB, N0447), 500 nM forward-index and reverse-payload primers, 0.2$\times$ EvaGreen (BIOTIUM, 31000-T), 1\% rAlbumin (NEB, B9200), 1\% Q5\textregistered{} HotStart High-Fidelity DNA polymerase (NEB, M0493) in a total volume of 20 $\upmu$L. Amplification was realized on CFX96 Touch Real-Time PCR Detection System using the following protocol: first denaturation step at 95$^{\circ}$C for 30 sec, 39 cycles of 95$^{\circ}$C 30 sec, 70$^{\circ}$C 30 sec and 72$^{\circ}$C 1 min and a final extension step at 72$^{\circ}$C for 5 min. To avoid heteroduplexes formation, PCR was followed in real time by fluorescent tracking and stopped at the end of the exponential phase by skipping step after 30 sec of extension at 72$^{\circ}$C. PCR product was purified using SPRIselect beads as previously mentioned using a 0.95$\times$ beads-to-sample ratio.\\
\\
{\it Estimation of the diversity.} Prior to sequencing the whole sample, a part was sequenced using Oxford Nanopore Technology (ONT) on  a Flongle flow cell to estimate key diversity. The remaining sample was kept at +4$^{\circ}$C until ready to use. Library preparation step was realized on amplified and purified dsDNA keys using the LSK-SQK114 ligation kit, with some modifications. The full protocol is available in  \textcolor{black}{Supplementary File 4a}. Libraries were sequenced on FLO-FLG114 flow cell.  Due to a sequencing crash, left-over libraries were loaded on a new flow cell. The reads  were filtered, clustered and the diversity estimated by fitting the cluster size distribution to a Poisson law.\\
\\
{\it Splitting amplified dsDNA keys.} The remaining sample was end-prepped following the supplier’s protocol provided in \textcolor{black}{Supplementary File 4b}. The sample was purified and was split in two parts of 31 $\upmu$L. The first part was kept in Paris for library preparation, while the other half was sent to Tokyo (LIMMS laboratory, Komaba Campus Tokyo University, Japan) at room temperature where it was kept at 4$^{\circ}$C until sequencing, roughly 1 month later. The adapter ligation step was prepared just before sequencing. \\
\\
{\it PromethION sequencing in Paris and Tokyo.}
A total of 130 fmol of libraries was loaded on PRO-MIN114 flow cell and sequenced on PromethION Solo 2 platform, generating 195.55M reads. The same protocol was applied in Tokyo, except that the left-over libraries ($\sim$ 50 fmol) were loaded on a second flow cell. First Tokyo run generated a total of 178.21M reads, while the second run generated 143.42M reads. After aligning and filtering, Alice and Bob's runs retained 146 033 874 and 201 264 655 reads, respectively.\\
\\%
{\bf Secure data sharing protocol: DNA key sample splitting at single copy stage and UMI tagging.}\\
\\
{\it Denaturation of dsDNA keys.} We first measured the melting temperature of dsDNA keys and found an experimental Tm of 74$^{\circ}$C. dsDNA keys were prepared as mentioned previously but using the following protocol: 25$^{\circ}$C for 10 sec and signal acquisition, 85$^{\circ}$C for 1 min, 84$^{\circ}$C to 60$^{\circ}$C at -0.5$^{\circ}$C per min, 60$^{\circ}$C for 40 sec and addition of 1\% Q5\textregistered{} Hot Start High-Fidelity DNA polymerase (NEB, M0493) at this point, 60$^{\circ}$C to 72$^{\circ}$C at +0.5$^{\circ}$C per min and a final extension at 72$^{\circ}$C for 10 min. The prepared dsDNA keys were bottlenecked via dilution and sampling to the targeted diversity and diluted in Milli-Q containing the double-stranded circular plasmid pUC19 (NEB, N3041), used as a carrier, at a final concentration of 4.7 nM. This sample was then denaturated by heating to 85$^{\circ}$C for 30 sec and cooling down to 24$^{\circ}$C with a rate of -3$^{\circ}$C per min. The denaturated sample was separated in two 2.9 $\upmu$L aliquots, for Alice and Bob. \\
\\
{\it Tagging the denaturated DNA keys with Unique Molecular Identifiers.} The forward and reverse UMI-primers were designed with 3 domains. From 5’ to 3’: a tail domain corresponding to the sequence of the external amplification primers (forward and reverse reamplification primers), used for library amplification, a short ${\rm N}_5$ UMI domain; a 3’ head identical to standard amplification primers (forward-index and reverse-payload primers, \textcolor{black}{Supplementary File 4}).
The DNA keys were submitted to two-cycles PCR using UMI-primers in a mix containing:  1$\times$ Q5\textregistered{} reaction buffer (NEB, M0493), 200 $\upmu$M dNTP (NEB, N0447), 200 nM forward and reverse UMI primers, 0.2$\times$ EvaGreen (BIOTIUM, 31000-T), 1\% rAlbumin (NEB, B9200), 1\% Q5\textregistered{} Hot Start High-Fidelity DNA polymerase (NEB, M0493). The following protocol was used: first denaturation at 98$^{\circ}$C for 15 sec, 2 cycles: 98$^{\circ}$C 10 sec, 70$^{\circ}$C for 30 sec, 72$^{\circ}$C for 1 min and a final extension step at 72$^{\circ}$C for 30 sec. Although Q5 enzyme requires a final concentration of primers of 500 nM, optimization had shown that the efficiency of this PCR does not deteriorate down to 200 nM of each primers.\\
\\
Excess UMI primers were then enzymatically digested. ExonucleaseI thermolabile enzyme (NEB, M0568) was diluted 6 times as follow: 6.3 $\upmu$L Milli-Q, 2 $\upmu$L Q5 reaction buffer 5X and 1.68 $\upmu$L exonucleaseI thermolabile. One $\upmu$L of this solution was added to the PCR tube and incubated at 20$^{\circ}$C for 10 min. After the reaction, the enzyme was deactivated for 1 min at 80$^{\circ}$C. To minimize pipetting and avoid losing molecules on surfaces, these reactions were conducted in the same PCR tube.\\
\\
{\it Amplifying UMI-tagged DNA keys for sequencing.} A PCR using external reamplification primers were realized with 1$\times$ Q5 reaction buffer (NEB, M0493), 200 $\upmu$M dNTP (NEB, N0447), 500 nM forward and reverse reamplification primers (\textcolor{black}{Supplementary File 5}), 0.2$\times$ EvaGreen (BIOTIUM, 31000-T), 1\% rAlbumin (NEB, B9200), 1\% Q5\textregistered{} Hot Start High-Fidelity DNA polymerase (NEB, M0493). The PCR was run with a first denaturation step at 98$^{\circ}$C for 15 sec, 39 cycles as follow: 98$^{\circ}$C for 10 sec, 65$^{\circ}$C for 30 sec and 72$^{\circ}$C for 1 min; and a last extension step at 72$^{\circ}$C for 5 min. The PCR was followed in real time by fluorescent tracking and stopped at the end of the exponential phase by skipping step after 30 sec of extension at 72$^{\circ}$C. Final extension was performed for 5 min at 72$^{\circ}$C. Lastly, PCR products were purified using SPRIselect magnetic beads using a beads-to-sample ratio of 0.95x and 85\% EtOH.\\
\\
{\it Attack scenario 1: Eve steals part of the message without replacement.} First, four single copy DNA pad pairs with diversity 2 million were created as mentioned above (See Denaturation of dsDNA keys). Four stealing fractions were tested: 0\% (no-stealing control), 10\%, 50\% and 90\%.  Eve sampled the corresponding volume (0.29 $\upmu$L, 1.95 $\upmu$L and 2.61 $\upmu$L) from Bob’s pad and replaced it with the same amount of Milli-Q water. \\
\\
Eve’s amplified the stolen sample via a standard DNA key amplification (See Amplifying dsDNA keys) but using the following protocol: first denaturation step at 98$^{\circ}$C for 10 sec, 39 cycles of 98$^{\circ}$C 15 sec, 70$^{\circ}$C 30 sec and 72$^{\circ}$C 1 min and a final extension step at 72$^{\circ}$C for 5 min. This protocol is later referred as “Eve Stealing Protocol” (ESP). She then prepared the sample for sequencing without UMI tagging.\\
\\
Alice and Bob tagged their denatured and split DNA keys with UMI and amplified their UMI keys as mentioned previously (See UMI-tagging and amplifying UMI-tagged sections).
ONT libraries were prepared on stolen amplified keys as well as Alice and Bob keys using the SQK-NBD114-24 library preparation kit with native barcoding as mentioned in \textcolor{black}{Supplementary File 4c.} Libraries (60 fmol) were loaded on a promethION flow cell and run on a PromethION Solo 2 platform. \\
\\
{\it Attack scenario 2: Eve steals Bob’s pads, amplifies it by PCR, splits and replaces.} First, four single copy DNA pad pairs with diversity 2 million were created as above. Eve took the entirety of Bob’s pads and performed PCR using the ESP protocol, butlimiting the cyle number to 1, 2 or 10 PCR cycles in order to adjust the copy number. One of Bob’s pads was left untouched as a control.\\
\\
To avoid being uncovered by left-over primers, Eve then degraded her primers using exonucleaseI thermolabile/6 following the protocol mentioned in {\it Tagging denaturated DNA keys with Unique Molecular Identifiers}. Then, she sampled her PCR product to collect 2 million DNA strands for Bob, amounting to half of her sample in the 1 cycle case,$1/4$ of for 2 cycles and 1/1000 of for the 10 cycles case. Additionally, she adjusted the volume restituted to Bob to 2.9 $\upmu$L. Prior to sequencing, Eve performed as described above (See ESP of Attack 1 section).
On their side, Alice and Bob performed UMI-tagging and sequencing as previously described (See {\it UMI-tagging} and {\it amplifying UMI-tDNA} sections). A total of 130 fmol of libraries was loaded onto a PromethION flow cell. \\
\\
{\bf Basecalling and sequence processing for mask generation.}\\
\\
{\it Basecalling.} Raw data were acquired using live basecalling except for the run of Tokyo. Basecalling for all experiments was performed using dorado v .1.1.1+3c7eef9 with the following model: dna\_r10.4.1\_e8.2\_400bps\_sup@v5.0.0 or v5.2.0. In the case of pooled sequencing, the different experiments were tagged with the native barcoding kit SQK-NBD114-24, and live demultiplexed using dorado demux  and the --{\it barcode-both-ends} parameter. \\
\\
All sequence processing followed the same pipeline, using custom code written in Mathematica or Python. The raw basecalls were filtered by median Qscore and length. (-) reads were converted to (+) reads and all were then aligned on a reference where the expected random regions were represented by N, allowing the aligning algorithm to match any bases at these positions without penalities. After aligning, the non-constant regions (including sequencing barcodes and random blocks -such as UMI, index blocks or payload blocks) were extracted from the sequence, along with their associated Qscores. Insertions were replaced with the first base at the corresponding position in alignment (and given the minimal value of the associated Qscores), to allow a dense tabular format. Qscores for deletions were computed as the min of the two nearest attributed Qscores.\\
\\
In the case of indexed sequencing ({\it i.e.}, using the native barcoding kit on ONT to pool multiple samples in the same sequencing run), we then extracted, for each barcode, only the reads with proper matching barcodes on both sides (allowing an edit distance of 3 for barcode attribution).\\
\\
{\it Clustering and consensus calling.} We then performed clustering of all reads according to their index and payload blocks (and ignoring UMIs in case UMIs were present) via an iterative process. The median Qscore $M_q$ of each block, averaged over all reads, was computed and ordered. We then selected the 6 blocks with the highest $M_q$  and concatenated the corresponding sequences to generate an $i_1$ read identifier (of length 30 nt). We then grouped the reads by perfect $i_1$ match. Within each group, we extracted the list of sequences for the second-best group of 6 blocks, and computed a consensus via simple majority voting, to generate $i_1$. If a group contained only one read, the raw sequence was used as $i_2$. We then iterated the grouping, combining all groups that had the same $i_2$. The process was repeated until all blocks had been used.\\
\\
For example, the number of clusters and the number of clusters containing more than one read along the iterative process for Bob’s sequencing shown in Fig.2 in the main text, is given in Supplementary Table S6.\\
\\
We then retained only the clusters with more than one read and computed a complete consensus, while also estimating the error probability (as a consensus Qscore) at each position. For each position, the consensus base was selected by weighted majority voting, where each weight was the Qscore provided by the basecaller for that base. Indeed, as the Qscore are defined on a logarithmic scale, the most likely base is the one with the highest total Qscore. Its consensus Qscore can then be approximated as the sum of all Qscores associated with the winning base, minus the sum of all Qscores associated with non-majority bases, minus a penalty of 4.8 per non majority base (\textcolor{black}{Supplementary note 8 for a derivation}).\\
\\
{\it Consensus filtering.} Once all consensuses were computed, the data was organized in an index/payload format by fusing blocks 8-14 as an index and 15-28 as a payload (that is, 6 blocks originating from the index strand as index, and all blocks from the payload strand as payload). This design offered an indexing capacity of $4^{30} \sim 10^{18}$ sequences. The blocks 1-7 were reserved for error estimation. The consensuses were then filtered, retaining only those with a minimum Qscore in the payload above 30. To filter out PCR errors (which are expected to generate a point-wise defect in the consensus per-base quality scores), we also computed the min value of the Qscores normalized by the block median and filtered out consensuses containing at least one value below 0.5. Finally we checked the unicity of each index and discarded the small fraction ($\sim$1/10000) of consensuses with non-unique indices, which may originate from PCR artifacts.\\
\\
{\it Synchronization and mask generation.} Bob sent the full list of his indices to Alice. Alice computed the intersection between this list and her own set of indices. For the large sequencing presented in Fig.2, Alice  found the intersection of indices to represent $\approx 82$\% of her total and sent that intersection in a random ordering $O_r$ back to Bob. Both Alice and Bob organized their set of payloads according to the ordering $O_r$ (thus discarding the keys whose index are not in the intersection) and created a file with the corresponding ordered payloads. They then grouped the payload sequences in blocks of 5, applied 5PPD, and concatenated the result column-wise to obtain a single large binary string.
The true error rate can be computed by comparing the mask computed independently by Alice and Bob. We found that the error was slightly higher ($\sim$4 folds) in protocol without PCR amplification before sample partitioning. Since the sequencing depth we used was typically higher for the later runs, this higher error rate does not originate from the sequencing. We speculated that this mild desynchronization originates from the PCR errors occurring during the two parallel amplifications from single molecules.

\bibliography{Biblio} 
\bibliographystyle{ieeetr}
\section{Acknowledgements} %
Dr. A.G. passed away on April 2025 Before his passing, the late A.G. was deeply involved in conceptualizing and implementing the present studies.  This manuscript is based on a  preliminary outline that he produced. We are incredibly grateful for his contributions and would like to dedicate this work to his memory. The authors also thank Yannick Tauran, Kévin Ricard for their help in sequencing, Guillaume Gines for critical comments, Nicolas Clément, Masahiro Nomura, Bruno Le Pioufle, the Service pour la Science et la Technologie  and the press service of the French Embassy in Japan for his their kind assistance and support. We would like to thank Yuri Klebanov and Naoto Takayama from the IIS Tokyo Design Lab for the design of a DNA capsule. This work is funded by the ANR DNASec (ANR-24-CE39-3908-04), and the PEPR MoleculArXiv (ANR-22-PEXM-0002).

\section{Authors contributions} SJ performed the experiments and analysed the results with YR and VS. HG and MC performed the entropy analysis. EB performed preliminary experiments. VB and VS contributed to the experimental workflow. TP, SG performed additional statistical analysis. All authors participated in data acquisition, writing and critical revision of the manuscript. YR, PG, ML, SHK, GC and AG were responsible for, conceptualization, supervision and funding acquisition. All authors (excepted AG) read and approved the final version of the manuscript.

\section{Declaration of interest}
 A patent reporting some of the methods  was filed by the CENTRE NATIONAL DE LA RECHERCHE SCIENTIFIQUE, INSTITUT MINES TELECOM, UNIVERSITE DE LIMOGES and \'ECOLE SUP\'ERIEURE DE PHYSIQUE ET DE CHIMIE INDUSTRIELLES DE LA VILLE DE PARIS.

\end{document}